\def\Rb{{I\!\! R}}
\def\Cb{\ \hbox{\vrule width 0.6pt height 6pt depth 0pt
                      \hskip -3.5 pt} C}
\documentstyle[12pt]{article}
\begin{document}
\parskip=3pt
\parindent=18pt
\baselineskip=20pt
\setcounter{page}{1}
\centerline{\Large\bf BRST Structures and Symplectic Geometry}
\vspace{2ex}
\centerline{\Large\bf on a Class of Supermanifolds}
\vspace{6ex}
\centerline{\large{\sf Sergio Albeverio} ~~~and~~~ {\sf Shao-Ming Fei}
\footnote{Alexander von Humboldt fellow.\\
\hspace{5mm}On leave from Institute of Physics, Chinese Academy of Sciences,
Beijing}}
\vspace{4ex}
\centerline{\sf Institute of Mathematics, Ruhr-University Bochum,
D-44780 Bochum, Germany\\}
\vspace{6.5ex}

\centerline{\large\bf   Abstract}
\begin{center}
\begin{minipage}{5in}
\vspace{3ex}
By investigating the symplectic geometry and geometric quantization on
a class of supermanifolds, we exhibit BRST structures for a certain kind of
algebras. We discuss the undeformed and q-deformed cases in the classical
as well as in the quantum cases.
\end{minipage}
\end{center}
\newpage
\bigskip
Quantum groups and quantum algebras \cite{qg} play an important
role in many physical problems such as exactly soluble models
in statistical mechanics, conformal field theory, integrable model field
theories \cite{app}. This is due to their rich mathematical structures and
geometric properties. Efforts have
also been made to explore possible applications to gauge theory
in terms of non-commutative geometry \cite{gauge}.
As the form of a q-gauge invariant theory should
be essentially determined by the constraint structure, the generalization
of the geometric structure of BRST invariance might provide
insight into possible applications of quantum groups to gauge
fields. The construction of the Balatin-Frandkin-Vilkovisky
BRST charge related to the $SU_q(2)$ algebra has also been discussed
from this point of view \cite{brstq}.

In fact it is also ~worthwhile to study the BRST constructions for q-deformed
algebras for general systems of non-first class constraints. In this letter
by investigating the symplectic geometry and geometric quantization on
a class of special supermanifolds we exhibit BRST structures
for a class of algebras including $SU_q(2)$, either undeformed or
q-deformed with three algebraic elements. We discuss the classical
as well as the quantum cases, as the q-deformation and
$\hbar$-quantization are different in principle [5,6].

It is well known that for a finite dimensional first
class constrained Hamitonian system, the constraints
$K_a$, $a=1,...,n$, satisfy Poisson algebra relations
\begin{equation}\label{01}
[K_a,K_b]_{P.B.}=f_{ab}^c K_c~,
\end{equation}
with $f^{a}_{bc}$ real coefficients and $[\,,\,]$ standing for
the Poisson bracket. The BRST charge $Q$ is given by \cite{brst}
\begin{equation}\label{02}
Q=C^a K_a +\frac{1}{2}f_{bc}^a\bar{C}_b C^c~,
\end{equation}
where $C^a$ and $\bar{C}_b$ are anticommuting grassmann quantities with
grassmann parities $1$ and $-1$ respectively and such that
$(C^a)^2=0$, $(\bar{C}_b)^2=0$, $C^a\bar{C}_b=-\bar{C}_b C^a$
and $[C_a,\bar{C}^b]_{P.B.}=\delta_{a}^{~b}$
($\delta_{a}^{~b}$ being the Kronecker's symbol). It is easy to prove that
the BRST charge $Q$ is nilpotent $[Q,Q]_{P.B.}=0$ and satisfies
\begin{equation}\label{04}
[Q,\bar{C}^a]_{P.B.}\stackrel{def.}{=}\tilde{K}_a =K_a
+f_{ab}^c \bar{C}_c C^b~,~~~~
[\tilde{K}_a,\tilde{K}_b]_{P.B.}=f_{ab}^c \tilde{K}_c~.
\end{equation}
\begin{equation}\label{05}
[\tilde{K}_a,\bar{C}_b]_{P.B.}=f_{ab}^c \bar{C}_c~,~~~
[Q,\tilde{K}_a]_{P.B.}=0~.
\end{equation}
\begin{equation}\label{06}
[Q,C^a]_{P.B.}+\frac{1}{2}f_{bc}^a C^b C^c=0~.
\end{equation}
Equation (\ref{06}) is the Maurer-Cartan equation related to the BRST
cohomology.

To give a systematic description of the BRST structures for general
constrained Hamiltonian systems with three constraints,
we consider the supermanifold $M=M_{+}\times M_{0}\times M_{-}$, where
$M_0$ is an usual commutative differentiable manifold
with even dimensions, $M_+$
and $M_-$ are the anticommuting parts of $M$ with grassmann
parity $+1$ resp. $-1$
(these parts correspond to ``ghost'' and ``antighost'' respectively
in the BRST formalism).
We will use freely notions and notations from supermanifold theory
as described in \cite{super},
here we limit ourselves to a brief summary. The tangent space
$T_{p}(M)$ of the supermanifold $M$ at
point $p$ is spanned by the local supervector basis
$_{i}e=\left(\frac{\stackrel{\rightarrow}{\partial}}{\partial x^{i}}
\right)_{p}$ or
$e_{i}=\left(\frac{\stackrel{\leftarrow}{\partial}}{\partial x^{i}}
\right)_{p}$,
satisfying the relations $_{i}e=(-1)^{i}e_{i}$,
where $x^{i}$ is the local coordinate
and the exponent of $(-1)$ is the grassmann parity of $x^{i}$.
The arrow represents the acting direction of the partial derivatives.
The grassmann
parity takes values $0$ or $\pm 1$ with respect to commuting
(``c-type'') variables
and anticommuting (``a-type'')
variables respectively. A tangent vector ${\bf X}$
on $M$ may be expressed as
${\bf X}=X^{i}~_{i}e=(-1)^{i(X+i)}\, _i e X^{i}=e_i\,^iX $,
where $^{i}X\stackrel{def.}{=}(-1)^{iX}X^{i}$ is the $i$-th coordinate of the
tangent space and $X$ in the exponent of $(-1)$ is the
grassmann parity of $X$.

The local basis for the dual space $T_{p}^{\star}(M)$ is
$e^i =\,^i e =dx^i$ satisfying
$<\,_i e,e^j>=\,_i \delta^j$ and $_i \delta^j=(-1)^i \delta_i^{~j}
=(-1)^{i+ij}\delta_{~i}^{j}$.
Similarly a local dual vector {\bf V} takes the form
${\bf V}=e^i\,_iV=(-1)^{i(i+V)}\,_i Ve^i =V_i\, ^i e$,
where $V_i\stackrel{def.}{=} (-1)^{i(V+1)}\, _i V$.

As $M_0$ is even dimensional, a real, c-type, super-antisymmetric and closed
two form $\omega$ can be defined. We have thus $d\omega=0$, $\omega$ is
a symplectic
form on the supermanifold $M$, and is non-degenerate, i.e., ${\bf X}=0$
if ${\bf X}\rfloor\omega=0$, where ${\bf X}$ is the supervector field on $M$
and $\rfloor$ denotes the left inner product defined by $({\bf X}\rfloor
\omega)({\bf Y})=\omega({\bf X},{\bf Y})$ for any two vectors ${\bf X}$
and ${\bf Y}$ on $M$. Locally,
\begin{equation}\label{6}
\omega=\displaystyle\frac{1}{2}dx^i\, _i\omega_j dx^j~,
\end{equation}
where $_i\omega_j$ has
the supersymmetric property, $_i\omega_j=-(-1)^{i+j+ij}\! _j\omega_i$ and
the non degeneracy property, $sdet(_i\omega_j)\neq 0$.
Here $~sdet$ stands for superdeterminant.

The canonical transformations are $\omega$-preserving diffeomorphisms of $M$
onto itself. A vector $X$ on $M$ corresponds to an infinitesimal
canonical transformation if and only if the Lie derivative of $\omega$ with
respect to ${\bf X}$ vanishes,
\begin{equation}\label{7}
{\cal L}_{{\bf X}}\omega={\bf X}\rfloor d\,\omega+d\,({\bf X}\rfloor\omega)=0~.
\end{equation}
A vector field ${\bf X}$ satisfying (\ref{7}) is said to be a Hamiltonian
vector field on the supermanifold $M$.
Let ${\cal F}(M)$ denotes the set of differentiable functions
on the supermanifold $M$. It
can be proved following \cite{wood} that for any function $f\in {\cal F}(M)$,
there exits a unique Hamiltonian vector field ${\bf X}_{f}$ satisfying
\begin{equation}\label{8}
{\bf X}_{f}\rfloor\omega=-d\,f~.
\end{equation}
For $f,g\in {\cal F}(M)$ the super Poisson bracket $[f,g]_{P.B.}$
is defined by
\begin{equation}\label{10}
[f,g]_{P.B.}=-\omega({\bf X}_f,{\bf X}_g)=\omega({\bf X}_g,{\bf X}_f)
=-{\bf X}_f g=(-1)^{fg}{\bf X}_g f~.
\end{equation}
It has the following properties
$$
\begin{array}{l}
[f,g]_{P.B.}=-(-1)^{fg}[f,g]_{P.B.}\\[3mm]
[f,[g,h]_{P.B.}]_{P.B.}
+(-1)^{f(g+h)}[g,[h,f]_{P.B.}]_{P.B.}
+(-1)^{h(f+g)}[h,[f,g]_{P.B.}]_{P.B.}=0~.
\end{array}
$$
For $f\in {\cal F}(M)$ the related Hamitonian vector field ${\bf X}_f$
has locally the form
\begin{equation}\label{11}
{\bf X}_f=f \displaystyle\frac{\stackrel{\leftarrow}{\partial}}
{\partial x^{i}}
\,^i\omega^j \displaystyle\frac{\stackrel{\rightarrow}{\partial}}{\partial
x^{j}}~.
\end{equation}
Hence from (\ref{10}) we have
\begin{equation}\label{12}
[f,g]_{P.B.}=-
f \displaystyle\frac{\stackrel{\leftarrow}{\partial}}{\partial x^{i}}
\,^i\omega^j \displaystyle\frac{\stackrel{\rightarrow}{\partial}}
{\partial x^{j}}g~,
\end{equation}
where $(^i\omega^j)$ is the inverse of $(_i\omega_j)$ (which
exists since by assumption $\omega$ is non degenerate).

Now we assume the manifold $M_0$ to be a 2-dimensional smooth manifold
defined by
\begin{equation}\label{13}
F(S_1,S_2,S_3)=0
\end{equation}
for some continuously differentiable function $F$: $\Rb^3\to \Cb$,
where $S_i$, $i=1,2,3$ are the dynamical variables on phase space, i.e., the
constraints of the Hamitonian system (i.e. we have the case
considered in (\ref{01}) with $n=3$). We recall that the
a-type coordinates $C^a$ resp. $\bar{C}_b$ of the
manifolds $M_+$ resp. $M_-$ with grassmann parity $+1$
resp. $-1$ satisfy (as in (\ref{01})):
\begin{equation}\label{14}
(C^a)^2=0~,~~~(\bar{C}_b)^2=0~,~~~C^a\bar{C}_b=-\bar{C}_b C^a~,~~~a,b=1,2,3~.
\end{equation}

{\sf Proposition 1.} The symplectic form $\omega$
on the supermanifold $M$ is given by
\begin{equation}\label{16}
\omega=\frac{1}{2}\sum_{ijk}\epsilon_{ijk}B_i dS_{j}\wedge dS_{k}+
\sum_a d\bar{C}_a\wedge dC^a~,
\end{equation}
where $B_i$ are differentiable functions of $S_i$ satisfying
\begin{equation}\label{17}
\sum_{i=1}^3 B_i\displaystyle\frac{\partial F}{\partial S_i}
=-\frac{1}{\alpha}~,
\end{equation}
for some real constant $\alpha$.
And for $f\in{\cal F}(M)$, the Hamiltonian vector field $X_f$ is
given by
\begin{equation}\label{18}
{\bf X}_f=
\alpha\sum_{ijk}\epsilon_{ijk}
\displaystyle\frac{\partial f}{\partial S_i}\displaystyle\frac{\partial
F}{\partial S_j}
\displaystyle\frac{\partial}{\partial S_k}+
f\displaystyle\frac{\stackrel{\leftarrow}{\partial}}{\partial C^a}
\displaystyle\frac{\stackrel{\rightarrow}{\partial}}{\partial \bar{C}_a}+
f\displaystyle\frac{\stackrel{\leftarrow}{\partial}}{\partial \bar{C}_a}
\displaystyle\frac{\stackrel{\rightarrow}{\partial}}{\partial C^a}~.
\end{equation}

{\sf [Proof].} The proof is straightforward by showing
${\bf X}_f\rfloor\omega=-df~,~~~~\forall f\in{\cal F}(M)$ in terms of
relations (\ref{13}) and (\ref{17}). It is also obvious that the
right hand side of (\ref{16}) gives a closed form since
$M_0$ is here a 2-dimensional manifold. $\P$

{}From the definition of the Poisson bracket (\ref{10}) and formula (\ref{18}),
the Poisson bracket for $f,g\in {\cal F}(M)$ is then given by
\begin{equation}\label{19}
[f,g]_{P.B.}=-\left(
\alpha\sum_{ijk}\epsilon_{ijk}
\displaystyle\frac{\partial f}{\partial S_i}\displaystyle\frac{\partial
F}{\partial S_j}
\displaystyle\frac{\partial g}{\partial S_k}+
f\displaystyle\frac{\stackrel{\leftarrow}{\partial}}{\partial C^a}
\displaystyle\frac{\stackrel{\rightarrow}{\partial}}{\partial \bar{C}_a}g+
f\displaystyle\frac{\stackrel{\leftarrow}{\partial}}{\partial \bar{C}_a}
\displaystyle\frac{\stackrel{\rightarrow}{\partial}}{\partial C^a}g\right)~.
\end{equation}
We remark that this Poisson bracket is uniquely given by the
manifold up to an algebraic isomorphism. It does not depend on the form
of the symplectic form $\omega$.

Formula (\ref{19}) gives rise to the classical Poisson bracket relations
of $\bar{C}_a$ and $C^b$,
\begin{equation}\label{19.5}
[C_a,\bar{C}^b]_{P.B.}=\delta_{a}^{~b}
\end{equation}
with all other Poisson brackets being zero.

{}From formulas (\ref{18}) and (\ref{19}) we easily have
\begin{equation}\label{20}
{\bf X}_{S_i}=
\alpha\sum_{ijk}\epsilon_{ijk}\displaystyle\frac{\partial F}{\partial S_j}
\displaystyle\frac{\partial}{\partial S_k}
\end{equation}
and
\begin{equation}\label{21}
[S_i,S_j]_{P.B.}=
\alpha\sum_{k}\epsilon_{ijk}\displaystyle\frac{\partial F}{\partial S_k}~.
\end{equation}
Therefore $S_i$, $i=1,2,3$ constitute a Poisson algebra. For
different values of the constant
$\alpha$ the Poisson algebras are algebraic isomorphic.
Henceforth for simplicity we will take
$\alpha$ to be $\frac{1}{2}$. Formula (\ref{21})
expresses the fact that the 2-dimensional manifold given by
(\ref{13}) is related to a
certain algebra. When $F(S_1,S_2,S_3)$ is a quadratic function of $S_i$,
then it gives rise to a linear algebra, the first class
constraint of the Hamiltonian system.

To investigate the BRST cohomology for both undeformed and q-deformed
cases in terms of symplectic geometry and geometric quantization, we
take the manifold $M_0$ defined by $F$ to be of the form
\begin{equation}\label{22}
F=S_1^2 +S_2^2+ G(S_3)-const.=0~,
\end{equation}
where $G$ is a continuously differentiable function. $const$
represents the Casimir invariant after geometric quantization.
{}From formula (\ref{21}) we deduce the
Poisson algebraic relations with respect to
the manifold (\ref{22}) (``classical constraint algebra''):
\begin{equation}\label{23}
[S_1,S_2]_{P.B.}=\displaystyle\displaystyle\frac{\partial G(S_3)}
{2\partial S_3}~,~~~
[S_2,S_3]_{P.B.}=S_1~,~~~[S_3,S_1]_{P.B.}=S_2~.
\end{equation}
This can be rewritten in the form
\begin{equation}\label{24}
[S_+,S_-]_{P.B.}=-i\displaystyle\frac{\partial G(S_3)}{\partial S_3}~,~~~~
[S_3,S_{\pm}]_{P.B.}=\mp iS_{\pm}~,
\end{equation}
where $S_{\pm}=S_1\pm iS_2$.
We set $C^{\pm}=\displaystyle\frac{1}{2}(C^1 \mp i C^2)$. From (\ref{19.5})
we have then
\begin{equation}\label{24.5}
[C^\pm,\bar{C}_\pm]_{P.B.}=\displaystyle\frac{1}{2}~,~~~[C^3,\bar{C}_3]_{P.B.}=1~.
\end{equation}

{\sf Proposition 2.} The BRST charge associated with the
algebra (\ref{23}) has the following general form
\begin{equation}\label{25}
\begin{array}{rcl}
Q&=&S_+ C^+ + S_- C^- +A_1  C^3 + A_2  \bar{C}_- C^- C^3+
A_3  \bar{C}_+ C^+ C^3 +
A_4  \bar{C}_3 C^+ C^- \\[3mm]
&&+ A_5  \bar{C}_+ \bar{C}_- C^+ C^- C^3 ~,
\end{array}
\end{equation}
where $A_l\equiv A_l(S_3)$, $l=1,...,5$ are continuously
differentiable functions. $Q$ is nilpotent (in the sense that
$[Q,Q]_{P.B.}=0$) if
\begin{equation}\label{28}
A_4 A_1=i\displaystyle\displaystyle\frac{\partial G(S_3)}{\partial S_3}~,~~~~
A_2=2i \displaystyle\frac{\partial A_1}{\partial S_3}~,~~~~
A_3=-A_2~,~~~~A_5=2i\displaystyle\frac{\partial A_2}{\partial S_3}~,
\end{equation}

{\sf [Proof].} From formula (\ref{19}) we have
\begin{equation}\label{26}
[S_-,A_l]_{P.B.}=-iS_- \displaystyle\frac{\partial A_l}{\partial S_3}~,~~~~
[S_+,A_l]_{P.B.}=iS_+ \displaystyle\frac{\partial A_l}{\partial
S_3}~.
\end{equation}
Let $Q$ be given by (\ref{25}). Then we have
\begin{equation}\label{27}
\begin{array}{rcl}
[Q,Q]_{P.B.}&=&2(-i\displaystyle\frac{\partial G(S_3)}{\partial S_3}C^+ C^-
+i S_+
\displaystyle\frac{\partial A_1}{\partial S_3}C^+ C^3
+i S_+ \displaystyle\frac{\partial A_2}
{\partial S_3}C^+
\bar{C}_- C^- C^3 \\[4mm]&&
+\displaystyle\frac{1}{2}S_+ A_3 C^+ C^3
+\displaystyle\frac{1}{2}S_+ A_5 \bar{C}_- C^+ C^- C^3
-i S_- \displaystyle\frac{\partial A_1}{\partial S_3}C^- C^3 \\[4mm]
&&+ \displaystyle\frac{1}{2}S_- A_2 C^- C^3
-iS_- \displaystyle\frac{\partial A_3}{\partial S_3}C^- \bar{C}_+ C^+ C^3
-\displaystyle\frac{1}{2}S_- A_5 \bar{C}_+ C^+ C^- C^3\\[4mm]&&
+A_1 A_4 C^+ C^-
+\displaystyle\frac{1}{2}A_2 A_4 C^- C^3 \bar{C}_3 C^+
+\displaystyle\frac{1}{2}A_3 A_4 C^+ C^3 \bar{C}_3 C^- )~.
\end{array}
\end{equation}
Hence we have $[Q,Q]_{P.B.}=0$ under condition (\ref{28}). $\P$

In addition, from the definition of Hamiltonian vector field
$X_Q$ associated with the BRST charge $Q$ we have
${\cal L}_{X_Q}\omega=0$. Hence the symplectic form
(\ref{16}) and the corresponding phase space are BRST invariant.

In this way we get a general BRST construction for all the constraint algebras
of the form
(\ref{24}). Each $(A_l)_{l=1,...,5}$ gives rise to a solution of
(\ref{28}) and hence to an
expression of BRST charge. Here are two simple kinds of solutions:

\begin{equation}\label{28.1}
{\bf i.}~~~A_1=\displaystyle\frac{\partial G(S_3)}{\partial S_3}~,~~~
A_2=-A_3=2i\displaystyle\frac{\partial^2 G(S_3)}{\partial^2 S_3}~,~~~
A_4=i~,~~~
A_5=-4\displaystyle\frac{\partial^3 G(S_3)}{\partial^3 S_3}
\end{equation}
which inserted in (\ref{25}) gives rise to the following expression of
$Q$ (denoted by $Q_1$):
\begin{equation}\label{28.2}
\begin{array}{rcl}
Q_1 &=&S_+ C^+ + S_- C^- +\displaystyle\frac{\partial G(S_3)}{\partial S_3}
C^3
+ 2i \displaystyle\frac{\partial^2 G(S_3)}{\partial^2 S_3} (\bar{C}_- C^- C^3
- \bar{C}_+ C^+ C^3)\\[4mm]
&&+i \bar{C}_3 C^+ C^- -4\displaystyle\frac{\partial^3 G(S_3)}{\partial^3 S_3}
\bar{C}_+ \bar{C}_- C^+ C^- C^3 ~.
\end{array}
\end{equation}

{\bf ii.} When $\displaystyle\frac{\partial G(S_3)}{\partial S_3}$ can be
written as
$2H(S_3)S_3$ for some function $H(S_3)$, another simple solution
of (\ref{28}) is
\begin{equation}\label{28.3}
A_1=S_3~,~~~
A_2=-A_3=2i~,~~~
A_4=2iH(S_3)~,~~~
A_5=0~.
\end{equation}
The corresponding $Q$ from (\ref{25}) is:
\begin{equation}\label{28.4}
Q_2 =S_+ C^+ + S_- C^- +S_3 C^3
+ 2i \bar{C}_- C^- C^3
-2i \bar{C}_+ C^+ C^3 +
2i H(S_3)\bar{C}_3 C^+ C^- ~.
\end{equation}

Let us now give two particular examples of functions $F$ entering
(\ref{22}). We take $G$ resp. $const$ in (\ref{22}) as given by
$G(S_3)=- \frac{\sinh(2 \gamma S_3)}{2\gamma \cosh\gamma}$
resp. $const=\frac{\sinh\gamma}{2\gamma\cosh\gamma}$ so that (\ref{22})
reaches
\begin{equation}\label{28.5}
S_1^2 +S_2^2 - \displaystyle\frac{\sinh(2 \gamma S_3)}{2\gamma \cosh\gamma}=
\displaystyle\frac{\sinh\gamma}{2\gamma\cosh\gamma}~.
\end{equation}
$\gamma=\ln q$ is a deformation parameter, $q>0$.

{\sf Proposition 3.} For the manifold $M_0$ given by (\ref{28.5})
the symplectic form
(\ref{16}) on the supermanifold $M$ is given by
\begin{equation}\label{28.6}
\begin{array}{rcl}
\omega&=&-\displaystyle\frac{\gamma\cosh\gamma}{\sinh\gamma}
\left(S_1\,dS_2\wedge dS_3+S_2\,dS_3\wedge dS_1\right)
+\displaystyle\frac{\sinh(2\gamma S_3)\cosh\gamma}{\cosh(2\gamma S_3)
\sinh\gamma}dS_1\wedge dS_2 \\[5mm]
&&+ \sum_{i=1}^3 d \bar{C}_i\wedge dC^i~.
\end{array}
\end{equation}

{\sf [Proof].} From proposition 1, we see that what we have to verify is
condition (\ref{17}).
Using (\ref{28.5}) we have $ 2B_1 S_1+ 2 B_2 S_ 2 -
\frac{\cosh(2\gamma S_3)}{\cosh\gamma}B_3=-2$. Substituting
the functions $B_i$
taken from (\ref{28.6}) we see that (\ref{17}) is satisfied. $\P$

{\sf Proposition 4.} The Poisson algebra related to the
manifold (\ref{28.5}) is
just the q-deformed simple harmonic oscillator algebra ${\cal H}_q (4)$,
\begin{equation}\label{28.7}
[S_+,S_-]_{P.B.}=i\displaystyle\frac{\cosh(2\gamma S_3)}{\cosh\gamma}~,~~~~
[S_3,S_\pm ]_{P.B.}=\mp iS_\pm~.
\end{equation}

{\sf [Proof].} The proof is straightforward by using formula
(\ref{19}). $\P$

Corresponding to the case (\ref{28.2}) we have the following
expression for the BRST charge
\begin{equation}\label{28.9}
\begin{array}{rcl}
Q&=&S_+ C^+ + S_- C^- - \displaystyle\frac{\cosh(2\gamma S_3)}{\cosh\gamma}
C^3
-i \displaystyle\frac{4\gamma\sinh(2\gamma S_3)}{\cosh\gamma}(\bar{C}_- C^- C^3
- \bar{C}_+ C^+ C^3)\\[4mm]
&&+i \bar{C}_3 C^+ C^-
+16\displaystyle\frac{\gamma^2 \cosh(2\gamma S_3)}{\cosh\gamma}
\bar{C}_+ \bar{C}_- C^+ C^- C^3 ~.
\end{array}
\end{equation}
We remark that in the limit $\gamma\to 0$
the manifold (\ref{28.5}) becomes the elliptic
paraboloid $S_1^2+S_2^2-S_3=\frac{1}{2}$
and the algebra (\ref{28.7}) becomes the ${\cal H}(4)$ algebra
(of the simple classical harmonic oscillator).

For another application we consider the symmetry in the Kepler
problem on a two
dimensional sphere. It is described \cite{higgs} by the algebra
generated by $R_1$, $R_2$ and $L$, defined by
\begin{equation}\label{28.10}
\begin{array}{rcl}
[R_1,R_2]_{P.B.}&=&\displaystyle\displaystyle\frac{1}{4}
((\lambda -8E)L+8\lambda L^3)
,\\[4mm]
[R_2,L]_{P.B.}&=&R_1~,~~~[L,R_1]_{P.B.}=R_2~,
\end{array}
\end{equation}
where $\lambda$ is the curvature of the sphere, $E$ is the energy eigenvalue,
$R_1$, $R_2$ are the Runge-Lenz vectors and $L$ is the 3-d component of the
angular momentum.

{\sf Proposition 5.} The symmetry algebra (\ref{28.10}) of the 2-dimensional
Kepler problem can be described by the symplectic
geometry on the supermanifold $M$ with $M_0$ given by
\begin{equation}\label{28.11}
R_1^2 +R_2^2 + \displaystyle\frac{1}{4}(\lambda-8E)L^2 +\lambda L^4 =C_K
\end{equation}
with supersymplectic form
\begin{equation}\label{28.12}
\omega=-\displaystyle\frac{1}{C_K}\left(
R_1\,dR_2\wedge dL+R_2\,dL\wedge dR_1
+\displaystyle\frac{(\lambda -8E)L+4\lambda L^3}{(\lambda -8E)+8\lambda L^2}
dR_1\wedge dR_2\right) + \sum_{i=1}^3 d \bar{C}_i\wedge dC^i~,
\end{equation}
where $C_K$ is a constant.

{\sf [Proof].} Comparing formulas (\ref{21}) and (\ref{28.10}) we have
$\frac{1}{2}\frac{\partial F}{\partial L}=\frac{1}{4}
((\lambda -8E)L+8\lambda L^3)$,
$\frac{1}{2}\frac{\partial F}{\partial R_2}=R_2$,
$\frac{1}{2}\frac{\partial F}{\partial R_1}=R_1$. Hence
$dF=d(R_1^2 +R_2^2 + \displaystyle\frac{1}{4}(\lambda-8E)L^2 +\lambda L^4))$,
which gives rise to the manifold given by (\ref{28.11}). The expression for
$\omega$ is then deduced using (\ref{16}). $\P$

The classical BRST charge given by the formula (\ref{28.2}) is then
\begin{equation}\label{28.13}
\begin{array}{rcl}
Q_1&=&
R_+ C^+ + R_- C^- + (\displaystyle\frac{1}{2} (\lambda -8E)L+4\lambda L^3) C^3
+i \bar{C}_3 C^+ C^-  - 96 \lambda L \bar{C}_+ \bar{C}_- C^+ C^- C^3\\[3mm]
&&+2i (\displaystyle\frac{1}{2} (\lambda -8E)+12\lambda L^2)
(\bar{C}_- C^- C^3 - \bar{C}_+ C^+ C^3)~.
\end{array}
\end{equation}

As here $\displaystyle\frac{\partial G(L)}{\partial L}=
(\displaystyle\frac{1}{2} (\lambda -8E)+4\lambda L^2)\cdot L$,
we have also another solution of type (\ref{28.3}) for $Q$, namely
\begin{equation}\label{28.14}
\begin{array}{rcl}
Q_2&=&
R_+ C^+ + R_- C^- +L_3 C^3 + 2i \bar{C}_- C^- C^3
-2i \bar{C}_+ C^+ C^3\\[4mm] &&  +
i (\displaystyle\frac{1}{2} (\lambda -8E)+4\lambda L^2)\bar{C}_3 C^+ C^- ~,
\end{array}
\end{equation}
where $R_\pm =R_1\pm i R_2$.

$\underline{ \sf Remark}$ By similar methods we can give the BRST
construction for other algebras given
by the supersymplectic geometry on $M$, with $M_0$ of the form (\ref{22}).

Now we discuss relations similar to the case of the first class constrained
systems (\ref{01}). Corresponding to the relations (\ref{04}) we define
$\tilde{S}_\pm =\tilde{S}_1 \pm i \tilde{S}_2 = [Q, \bar{C}_1 \pm i
\bar{C}_2]_{P.B.}$. Using the general form of BRST charge (\ref{25}) we have
\begin{equation}\label{29}
\begin{array}{l}
\tilde{S}_+ =[Q, 2\bar{C}_+]_{P.B.}=
S_+ - A_3 \bar{C}_+ C^3 - A_4 \bar{C}_3 C^- + A_5 \bar{C}_+ \bar{C}_- C^-
C^3\\[3mm]
\tilde{S}_- =[Q, 2\bar{C}_-]_{P.B.}=
S_- - A_2 \bar{C}_- C^3 + A_4 \bar{C}_3 C^+ - A_5 \bar{C}_+ \bar{C}_- C^+
C^3\\[3mm]
\tilde{S}_3 =[Q, \bar{C}_3]_{P.B.}=
A_1 + A_2 \bar{C}_- C^- + A_3 \bar{C}_+ C^+ + A_5 \bar{C}_+ \bar{C}_- C^+ C^-
\end{array}
\end{equation}

{\sf Proposition 6.}
The algebraic relations of $\tilde{S}_{\pm,3}$ are in general no longer
similar to the
ones in (\ref{04}). Nevertheless, in the case of the solution (\ref{28.3})
one has
\begin{equation}\label{31}
[\tilde{S}_+,\tilde{S}_-]_{P.B.}=-2iH(S_3)\tilde{S}_3
+iS_+ \displaystyle\frac{\partial A_4}{\partial S_3} \bar{C}_3 C^+
-iS_- \displaystyle\frac{\partial A_4}{\partial S_3}\bar{C}_3 C^-~,~~~
[\tilde{S}_3,\tilde{S}_{\pm}]_{P.B.}=\mp i\tilde{S}_{\pm}~,
\end{equation}

{\sf [Proof].} A direct calculation of
relations among $\tilde{S}_{\pm,3}$ gives
\begin{equation}\label{30}
\begin{array}{rcl}
[\tilde{S}_3, \tilde{S}_+ ]_{P.B.}&=&
-i S_+ \displaystyle\frac{\partial A_1}{\partial S_3}
- \displaystyle\frac{1}{2}A_3^2 \bar{C}_+ C^3
+ \displaystyle\frac{1}{2}A_2 A_4 \bar{C}_3 C^-
+A_3 A_5 \bar{C}_+ \bar{C}_- C^- C^3\\[4mm]
&&-iS_+ (\displaystyle\frac{\partial A_2}{\partial S_3}\bar{C}_- C^-
+\displaystyle\frac{\partial A_3}{\partial S_3} \bar{C}_+ C^+ )
-iS_+ \displaystyle\frac{\partial A_5}{\partial S_3}
\bar{C}_+ \bar{C}_- C^+ C^- \\[4mm]&&
-\displaystyle\frac{1}{2}A_4 A_5 \bar{C}_3 \bar{C}_+ C^+ C^-\\[4mm]

[\tilde{S}_3, \tilde{S}_- ]_{P.B.}&=&
i S_- \displaystyle\frac{\partial A_1}{\partial S_3}
- \displaystyle\frac{1}{2}A_3^2 \bar{C}_- C^3
- \displaystyle\frac{1}{2}A_3 A_4 \bar{C}_3 C^+
-A_2 A_5 \bar{C}_+ \bar{C}_- C^+ C^3\\[4mm]
&&+iS_- (\displaystyle\frac{\partial A_2}{\partial S_3}\bar{C}_- C^-
+\displaystyle\frac{\partial A_3}{\partial S_3} \bar{C}_+ C^+ )
+iS_- \displaystyle\frac{\partial A_5}{\partial S_3}\bar{C}_+
\bar{C}_- C^+ C^-\\[4mm]&&
-\displaystyle\frac{1}{2}A_4 A_5 \bar{C}_3 \bar{C}_- C^+ C^-\\[4mm]

[\tilde{S}_+, \tilde{S}_- ]_{P.B.}&=&
-i \displaystyle\frac{\partial G}{\partial S_3} - A_2 A_4 \bar{C}_-
C^- - A_3 A_4 \bar{C}_+ C^+
- 2 A_4 A_5 \bar{C}_+ \bar{C}_- C^+ C^-\\[4mm]&&
-i S_+ \displaystyle\frac{\partial A_2}{\partial S_3}\bar{C}_- C^3
-i S_- \displaystyle\frac{\partial A_3}{\partial S_3}\bar{C}_+ C^3
+iS_+ \displaystyle\frac{\partial A_4}{\partial S_3} \bar{C}_3 C^+
-iS_- \displaystyle\frac{\partial A_4}{\partial S_3}\bar{C}_3 C^-\\[4mm]
&&-iS_+ \displaystyle\frac{\partial A_5}{\partial S_3} \bar{C}_+ \bar{C}_- C^+
C^3
+iS_- \displaystyle\frac{\partial A_5}{\partial S_3} \bar{C}_+ \bar{C}_- C^-
C^3\\[4mm]&&
-\displaystyle\frac{1}{2} A_4 A_5(\bar{C}_3 \bar{C}_+ C^+ C^3
+\bar{C}_3 \bar{C}_- C^- C^3)
\end{array}
\end{equation}
Substituting (\ref{28.3}) into (\ref{30}) one gets (\ref{31}). $\P$

By using the algebraic relations (\ref{24}) and (\ref{24.5})
it is also easy to
verify relations similar to (\ref{05}) and (\ref{06})
when the solution (\ref{28.3}) is taken into account.

{\sf Proposition 7.} In the case described by (\ref{28.3}) we have
\begin{equation}\label{a1}
[\tilde{S}_a,\bar{C}_b]_{P.B.}=f_{ab}^c \bar{C}_c~,~~~
[Q,\tilde{S}_a]_{P.B.}=0~,
\end{equation}
\begin{equation}\label{a2}
[Q,C^a]_{P.B.}+\frac{1}{2}f_{bc}^a C^b C^c=0~,
\end{equation}
where $a,b,c=1,2,3$, $f^a_{bc}=-f^a_{cb}$, $f^3_{12}=H(S_3)$,
$f^1_{23}=f^2_{31}=1$, $\tilde{S}_1=\frac{1}{2}(\tilde{S}_+ +\tilde{S}_-)$,
$\tilde{S}_2=\frac{1}{2i}(\tilde{S}_+ -\tilde{S}_-)$. $\P$

Therefore when $\frac{\partial G(S_3)}{\partial S_3}$ can be analytically
written as $2H(S_3)S_3$, for the algebra (\ref{24}) there exits a kind of BRST
construction with Maurer-Cartan equations given by
(\ref{a2}) and relations
(\ref{a1}), similar to the formulas (\ref{06}) and (\ref{05})
in the first class constrained system.

Now we consider the quantization of above BRST systems. In terms of geometric
quantization, the physical system is quantized by constructing a
prequantization line bundle on the symplectic manifold $(M,\omega)$ such that
its connection one form is a symplectic potential and the section
curvature is $\omega$, and introducing a polarization \cite{wood}.
The quantum Hilbert space is defined to be the subspace of the product
bundle's section space which is covariantly constant along the chosen
polarization.
For BRST systems, the additional condition is that $\hat{Q}$
annihilates the physical Hilbert space, where $\hat{Q}$ is the quantum
operator associated with the classical BRST charge $Q$.
The quantization gives rise to a map
from classical observables $f\in {\cal F}(M)$ to quantum operators $\hat{f}$.
The expression of the quantum operator $\hat{Q}$
depends on its relation to the classical
phase space as expressed by $Q$,
and the selection of polarization. Here rather than discussing in details
a physical constrained system, we will only discuss the
algebraic construction of a quantum BRST system.

The role of the manifold (\ref{22}) is taken up in the quantum case by the
corresponding Casimir operator (for $SU_q(2)$ see \cite{fg}). The algebraic
relations (\ref{24})  and (\ref{24.5}) become quantum ones,
\begin{equation}\label{a3}
[\hat{S}_+,\hat{S}_-]=\displaystyle\frac{\partial G(S_3)}
{\partial S_3}\vert_{S_3\to \hat{S}_3}\stackrel{def.}{=}[[\hat{S_3}]]~,~~~~
[\hat{S}_3,\hat{S}_{\pm}]=\pm \hat{S}_{\pm}~,
\end{equation}
\begin{equation}\label{a4}
[\hat{C}^\pm,\hat{\bar{C}}_\pm]=\displaystyle\frac{i}{2}~,~~~
[\hat{C}^3,\hat{\bar{C}}_3]=i~,
\end{equation}
where $[\,,\,]$ represents the
supercommutator defined by $[A,B]=AB-(-1)^{AB}BA$.
For the Schr\"odinger polarization of the anticommuting part there are
explicit expressions, and $\hat{C}^a=C^a$, $\hat{\bar{C}}_a
=i\frac{\stackrel{\rightarrow}{\partial}}{\partial C^a}$, $a=1,2,3$,
see \cite{fgy}.

{\sf Proposition 8}. The quantum BRST charge given by
\begin{equation}\label{a5}
\begin{array}{rcl}
\hat{Q}&=&S_+ \hat{C}^+ + S_- \hat{C}^- +A_1(\hat{S}_3)  \hat{C}^3
+ A_2(\hat{S}_3)  \hat{\bar{C}}_- \hat{C}^- \hat{C}^3+
A_3(\hat{S}_3)  \hat{\bar{C}}_+ \hat{C}^+ \hat{C}^3 \\[3mm]
&&+A_4(\hat{S}_3)  \hat{\bar{C}}_3 \hat{C}^+ \hat{C}^- +
A_5(\hat{S}_3)  \hat{\bar{C}}_+ \hat{\bar{C}}_- \hat{C}^+ \hat{C}^- \hat{C}^3
{}~,
\end{array}
\end{equation}
is nilpotent in the sense that $[\hat{Q},\hat{Q}]=2\hat{Q}^2=0$. Here we have
\begin{equation}\label{a6}
\begin{array}{ll}
A_4(\hat{S}_3) A_1(\hat{S}_3)=i[[\hat{S}_3]]~,~~~~~~~&
A_2(\hat{S}_3)=2i (A_1(\hat{S}_3)-A_1(\hat{S}_3-1))\\[3mm]
A_3(\hat{S}_3)=2i (A_1(\hat{S}_3)-A_1(\hat{S}_3+1))~,&
A_5=4(2A_1(\hat{S}_3)-A_1(\hat{S}_3-1)-A_1(\hat{S}_3+1))~.
\end{array}
\end{equation}
{\sf [Proof].} The proof is straightforward by
using relations (\ref{a3}) and (\ref{a4}). $\P$

Corresponding to the classical cases (\ref{28.2}) resp. (\ref{28.4}), we
have two simple solutions of
(\ref{a6}) and two quantum expressions of $\hat{Q}$,
\begin{equation}\label{a7}
\begin{array}{ll}
{\bf i.} &A_1(\hat{S}_3)=[[\hat{S}_3]]~,~~~
A_2(\hat{S}_3)=2i ([[\hat{S}_3]]-[[\hat{S}_3-1]])~,~~~~
A_4(\hat{S}_3)=i~,\\[3mm]
&A_3(\hat{S}_3)=2i ([[\hat{S}_3]]-[[\hat{S}_3+1]])~,~~~~
A_5=4(2[[\hat{S}_3]]-[[\hat{S}_3-1]]-[[\hat{S}_3+1]])~.
\end{array}
\end{equation}
\begin{equation}\label{a8}
\begin{array}{rcl}
\hat{Q}_1&=&\hat{S}_+ \hat{C}^+ + \hat{S}_- \hat{C}^-
+[[\hat{S}_3]]  \hat{C}^3 +
2i ([[\hat{S}_3]]-[[\hat{S}_3-1]]) \hat{\bar{C}}_- \hat{C}^- \hat{C}^3\\[3mm]
&&+2i ([[\hat{S}_3]]-[[\hat{S}_3+1]])  \hat{\bar{C}}_+ \hat{C}^+ \hat{C}^3
+i  \hat{\bar{C}}_3 \hat{C}^+ \hat{C}^-\\[3mm]
&&+4(2[[\hat{S}_3]]-[[\hat{S}_3-1]]-[[\hat{S}_3+1]])
\hat{\bar{C}}_+ \hat{\bar{C}}_- \hat{C}^+ \hat{C}^- \hat{C}^3 ~.
\end{array}
\end{equation}

{\bf ii.} When $[[\hat{S}_3]]$ can be analytically written as
$2H(\hat{S}_3)\hat{S}_3$, we have
\begin{equation}\label{a9}
A_1(\hat{S}_3)=\hat{S}_3~,~~
A_2(\hat{S}_3)=-A_3(\hat{S}_3)=2i~,~~
A_4(\hat{S}_3)=2iH(\hat{S}_3)~,~~
A_5(\hat{S}_3)=0~,
\end{equation}
\begin{equation}\label{a10}
\hat{Q}_2=\hat{S}_+ \hat{C}^+ + \hat{S}_- \hat{C}^- + \hat{S}_3  \hat{C}^3
+2i  \hat{\bar{C}}_- \hat{C}^- \hat{C}^3
-2i  \hat{\bar{C}}_+ \hat{C}^+ \hat{C}^3
+2iH(\hat{S}_3)  \hat{\bar{C}}_3 \hat{C}^+ \hat{C}^- ~.
\end{equation}

For the more particular case of the
${\cal H}_q(4)$ algebra (~occurring in Prop. 4), (\ref{28.7}) becomes
$$
[\hat{S}_+,\hat{S}_-]=-\displaystyle\frac{\cosh(2\gamma
\hat{S}_3)}{\cosh\gamma}~,~~~~
[\hat{S}_3,\hat{S}_\pm ]=\pm \hat{S}_\pm~.
$$
Using the solution (\ref{a7}) we have the quantum BRST charge
$$
\begin{array}{rcl}
\hat{Q}&=&\hat{S}_+ \hat{C}^+ + \hat{S}_- \hat{C}^-
-\displaystyle\frac{\cosh(2\gamma \hat{S}_3)}{\cosh\gamma}  \hat{C}^3 +
\frac{2i}{\cosh\gamma} (\cosh 2\gamma(\hat{S}_3 -1)-\cosh 2\gamma \hat{S}_3)
\hat{\bar{C}}_- \hat{C}^- \hat{C}^3 \\[3mm]
&&+ \displaystyle\frac{2i}{\cosh\gamma}
(\cosh 2\gamma(\hat{S}_3 +1)-\cosh 2\gamma \hat{S}_3)
\hat{\bar{C}}_+ \hat{C}^+ \hat{C}^3
-i  \hat{\bar{C}}_3 \hat{C}^+ \hat{C}^- \\[3mm]
&&+\displaystyle\frac{8}{\cosh\gamma}
\cosh (2\gamma \hat{S}_3) (\cosh 2\gamma -1)
\hat{\bar{C}}_+ \hat{\bar{C}}_- \hat{C}^+ \hat{C}^- \hat{C}^3 ~.
\end{array}
$$

For the algebra (\ref{28.10}) of the Kepler problem on the 2-sphere,
the quantization gives rise to the commutation relations
$$
\begin{array}{rcl}
[\hat{R}_1,\hat{R}_2]&=&\displaystyle\frac{i}{4}
((\lambda -8E)\hat{L}+8\lambda \hat{L}^3)
,\\[4mm]
[\hat{R}_2,\hat{L}]&=&i\hat{R}_1~,~~~[\hat{L},\hat{R}_1]=i\hat{R}_2~,
\end{array}
$$
Corresponding to (\ref{a8}) and (\ref{a10}) we have
$$
\begin{array}{rcl}
\hat{Q}_1&=&\hat{R}_+ \hat{C}^+ + \hat{R}_- \hat{C}^-
+ \displaystyle\frac{1}{2}((\lambda-8E)\hat{L}+8\lambda
\hat{L}^3)\hat{C}^3\\[4mm]
&&+i(\lambda-8E +8\lambda(3\hat{L}^2-3\hat{L}+1))
\hat{\bar{C}}_- \hat{C}^- \hat{C}^3\\[4mm]
&&+i(8E-\lambda-8\lambda(3\hat{L}^2+3\hat{L}+1))
\hat{\bar{C}}_+ \hat{C}^+ \hat{C}^3\\[4mm]
&&+i\hat{\bar{C}}_3 \hat{C}^+ \hat{C}^-
-96\lambda \hat{L} \hat{\bar{C}}_+
\hat{\bar{C}}_- \hat{C}^+ \hat{C}^- \hat{C}^3 ~.\\[6mm]
\hat{Q}_2&=&\hat{R}_+ \hat{C}^+ + \hat{R}_- \hat{C}^- + S_3  \hat{C}^3
+2i  \hat{\bar{C}}_- \hat{C}^- \hat{C}^3
-2i  \hat{\bar{C}}_+ \hat{C}^+ \hat{C}^3 \\[3mm]
&&+\displaystyle\frac{i}{2}(\lambda -8E +8\lambda \hat{L}^2)  \hat{\bar{C}}_3
\hat{C}^+ \hat{C}^- ~,
\end{array}
$$
where $\hat{R}_\pm =\hat{R}_1 \pm i \hat{R}_2$.

By using the solution (\ref{a9}), we have

{\sf Proposition 9.}
$$
[\hat{\tilde{S}}_a,\hat{\bar{C}}_b]=f_{ab}^c \hat{\bar{C}}_c~,~~~
[\hat{Q},\hat{\tilde{S}}_a]=0~,
$$
$$
[\hat{Q},\hat{C}^a]+\frac{1}{2}f_{bc}^a \hat{C}^b \hat{C}^c=0~,
$$
where $a,b,c=1,2,3$, $f^{a}_{bc}=-f^a_{cb}$, $f^3_{12}=iH(\hat{S}_3)$,
$f^1_{23}=f^2_{31}=i$, $i\hat{\tilde{S}}_a=[\hat{Q}, \hat{\bar{C}}_a]$. $\P$

Summarizing, we have investigated the symplectic geometry on
a class of supermanifolds $M$ with $M_0$ defined by (\ref{22}).
The related BRST structures with classical constraint
algebra (\ref{24}) have been discussed in detail, as well as
the quantum BRST
structure with respect to the quantum algebra (\ref{a3}).
Two examples were given as applications. The BRST systems we discussed here
apply to a class of constrained systems with
both undeformed and q-deformed algebras, for instance,
$SU_q(2)$ and $SU_q(1,1)$, their related 2-dimensional manifolds
are just special examples of the manifold described by (\ref{22}),
see \cite{fg}.

In addition, we found that the expression of the BRST charge is not unique.
Other constructions from equations (\ref{28}) (classical case) resp.
(\ref{a6}) (quantum case) can also be
investigated. It would also be
interesting to consider the case where $q=e^\gamma$ is a root of unit.

\end{document}